\documentclass[11pt]{article}
\usepackage{amsmath,amssymb,amsthm,empheq}
\usepackage{amsfonts}
\usepackage{graphicx}
\usepackage{tikz}
\usepackage{url}
\usepackage{cite}
\usepackage{enumitem}

\begin{document}

\title{Solving break minimization problems \\ in mirrored double round-robin tournament \\ with QUBO solver
\thanks{
This work was supported by JSPS KAKENHI Grant Number JP20K04973. 
}
}

\author{Koichi Fujii
\thanks{Department of Industrial Engineering and Economics, 
   Tokyo Institute of Technology 
}
       \and
        Tomomi Matsui
\footnotemark[2]
}

\maketitle

\begin{abstract}
  The break minimization problem is
    a fundamental problem in sports scheduling.
  Recently, its quadratic unconstrained binary optimization (QUBO) formulation
    has been proposed,
    which has gained much interest with the rapidly growing field of quantum computing.
  In this paper, we demonstrate that
    the state-of-the-art QUBO solver outperforms
    the general mixed integer quadratic programming (MIQP) solver
    on break minimization problems in a mirrored double round-robin tournament.
  Moreover, we demonstrate that it still outperforms or is competitive
    even if we add practical constraints,
    such as consecutive constraints,
    to the break minimization problem.
\end{abstract}

\noindent
Keywords: sports scheduling, round robin tournament, QUBO
 
\section{Introduction}\label{sec:intro}

The construction of a suitable schedule for sports competitions is
  a crucial issue in sports scheduling.
The round-robin tournament is a competition adopted 
  in many professional sports,
  such as soccer and basketball.
This paper deals with a home-away assignment problem,
  which assigns home or away to each match of a double round-robin tournament schedule
  to minimize the number of breaks.

In recent years, Urdaneta, Yuan, and Sinqueria~\cite{urdaneta2018alternative} 
  formulated the break minimization problem in a double round-robin tournament 
  as a mixed integer quadratic programming (MIQP) with no constraint
  or a quadratic unconstrained binary optimization problem (QUBO).
They demonstrated the advantage of QUBO formulation
  against other formulations through numerical experiments.
Kuramata, Katsuki, and Nakata~\cite{kuramata2022solving} 
  successfully applied
  quantum annealing to the QUBO formulation,
  obtaining better solutions quickly.

QUBO has recently gained much attention owing to its relationship with quantum computing.
One of the best QUBO solvers was developed by Rehfeldt, Koch and Shinano~\cite{rehfeldt2023faster}.
They showed that their solver generally outperforms
  quantum techniques, including quantum annealing,
  although it is designed to prove optimality,
  contrary to heuristic approaches of quantum techniques.

In this paper,
  we demonstrate that the QUBO solver outperforms general MIQP solver
  for break minimization problems in a mirrored double round-robin tournament schedule.
We also extend the break minimization problem by adding consecutive constraints,
  which prohibit more than two consecutive home or away matches.
Formulating these extended problems as a QUBO model through penalization,
  we demonstrate that the QUBO solver can also outperform the MIQP solver.

The remainder of the paper is organized as follows.
Section~\ref{sec:breakmin} describes the break minimization problem in a double round-robin tournament schedule.
Section~\ref{sec:mathform} gives a mathematical programming formulation of the break minimization problem.
  We also show how to insert consecutive constraints into our formulations  
  of the break minimization problem.
Section~\ref{sec:comp} provides computational results 
  comparing the general MIQP and QUBO solvers
  for the break minimization problem.
It is shown that the QUBO solver outperforms or is competitive with
  the general MIQP solver,
  not only for the break minimization problem
  but also for problems with consecutive constraints.


\section{Break Minimization}\label{sec:breakmin}

In this section, we introduce round-robin tournament schedules,
  which frequently arise in many sports scheduling problems.

A {\it single round-robin tournament} (SRRT) is a simple sports competition,
  where each team plays against every other team once.
A {\it double round-robin tournament} (DRRT) is an extended version of an SRRT,
  where each team plays against all others twice.
In this paper, 
  we consider a DRRT schedule with the following properties:

\begin{itemize}
\item Each team plays one match in each \textit{slot}, the day when a match is held.
\item Each team has a home stadium, and each match is held 
  at the home stadium of one of the two playing teams.
  When a team plays a match at the home of the opponent, 
  we say that the team plays away.
\item Each team plays each other team exactly once at its home stadium and once away. 
\end{itemize}

A {\it mirrored DRRT} (MDRRT)
   is a DRRT with the same combination of matches 
   in the first and second halves.

\subsection{Preliminaries}
 
We use the following notations and symbols throughout this paper:

\begin{itemize}
  \item $2n$: number of teams, where $n \geq 2$, 
  \item $T = \{1,2, \ldots, 2n\}$: set of teams,
  \item $S = \{1,2, \ldots , 2(2n-1) \}$: set of slots.
\end{itemize}

A schedule of a DRRT 
  is described as a pair of a timetable and a home–away assignment defined below. 
A timetable $\tau$ of a DRRT is 
  a matrix whose rows and columns are indexed by $T$ and $S$, respectively.
An element $\tau (t,s)$ shows the opponent that plays against team $t$ at slot $s$.
A timetable $\tau$ (of a DRRT schedule) should meet the following conditions:
  (i)~a row of $\tau$ indexed by team $t \in T$ is obtained 
  by concatenating two permutations of $T\setminus \{t\}$,
  and (ii)~$\tau(\tau(t,s), s) = t \; (\forall (t,s)\in T \times S).$
In the case of an MDRRT, 
  the corresponding timetable satisfies that 
  each row of $\tau$ indexed by $t$ is a repetition of a mutual permutation; 
  i.e., $\tau(t,s) = \tau(t, s + 2n - 1)$ holds for each
  $ (t,s) \in T \times \{ 1 , \ldots , 2n - 1  \}$.
%
Table~\ref{table:timetable} shows a timetable for an MDRRT schedule.

\begin{table}[htbp]
  \begin{minipage}{.50\textwidth}
    \caption[short]{Timetable}
    \label{table:timetable}
    \begin{tabular}{|l|l|l|l|l|l|l|}
      \hline Slot & $\mathbf{1}$ & $\mathbf{2}$ & $\mathbf{3}$ & $\mathbf{4}$ & $\mathbf{5}$ & $\mathbf{6}$ \\
      \hline team 1 & 2 & 3 & 4 & 2 & 3 & 4 \\
      \hline team 2 & 1 & 4 & 3 & 1 & 4 & 3 \\
      \hline team 3 & 4 & 1 & 2 & 4 & 1 & 2 \\
      \hline team 4 & 3 & 2 & 1 & 3 & 2 & 1 \\
      \hline
    \end{tabular}
\end{minipage}
  \begin{minipage}{.50\textwidth}
    \caption[short]{Howe-away assignment}    
    \label{table:ha_assignment}
    \begin{tabular}{|l|l|l|l|l|l|l|}
    \hline Slot & $\mathbf{1}$ & $\mathbf{2}$ & $\mathbf{3}$ & $\mathbf{4}$ & $\mathbf{5}$ & $\mathbf{6}$ \\
    \hline team 1 & 1 & 0 & 1 & 0 & 1 & 0 \\
    \hline team 2 & 0 & 0 & 1 & 1 & 1 & 0 \\
    \hline team 3 & 1 & 1 & 0 & 0 & 0 & 1 \\
    \hline team 4 & 0 & 1 & 0 & 1 & 0 & 1 \\
    \hline
  \end{tabular}
 \end{minipage}  
\end{table}

Given a timetable $\tau$ of a DRRT,
  we introduce a set defined by 
\[
  \Xi(\tau) = \{ (t_1, t_2, s_1, s_2) \in T^2 \times S^2  
  \mid t_1 < t_2, s_1 < s_2, t_2=\tau(t_1, s_1) =\tau(t_1, s_2)\}.
\]
We denote $\Xi(\tau)$ by $\Xi$ if there is no ambiguity.
Each element $(t_1, t_2, s_1, s_2) \in \Xi$ satisfies the following properties:
\begin{itemize}
  \item $t_1, t_2$: a pair of teams with $t_1 < t_2$.
  \item $s_1$: the slot of the first match between $t_1$ and $t_2$.
  \item $s_2$: the slot of the second match between $t_1$ and $t_2$.
\end{itemize}
Clearly, the cardinality of the set $\Xi$ is equal to $n(2n-1)$.
For an MDRRT schedule, 
  $s_2 = s_1 + 2n-1$ holds for each $(t_1, t_2, s_1, s_2) \in \Xi$.
The set $\Xi$ corresponding to the timetable in Table~\ref{table:timetable} becomes
\[
  \Xi = \{(1, 2, 1, 4), (1, 3, 2, 5), (1, 4, 3, 6), (2, 3, 3, 6), (2, 4, 2, 5), (3, 4, 1, 4)\}.
\]

A team is said to be at home in slot $s$ if the team plays a match 
  at its home stadium in $s$; otherwise away in $s$.
A {\it home-away assignment} (HA-assignment) is a matrix $Y$
  whose rows are indexed by $T$, 
  and columns are indexed by $S$.
An element $y(t,s)$ of $Y$ is equal to 1
  if team $t$ plays a match in $s$ at its home stadium; 
  otherwise it is 0.
Tables~\ref{table:timetable} and~\ref{table:ha_assignment} 
  together show an example of an MDRRT schedule.
For a given timetable $\tau$, 
  an HA-assignment $Y$ is {\em consistent} with $\tau$ when the following holds:
\begin{equation} \label{eqn:xi}
\forall (t_1, t_2, s_1, s_2) \in \Xi (\tau), 
  \left(
    \begin{array}{cc}
      y(t_1, s_1) & y(t_1, s_2) \\
      y(t_2, s_1) & y(t_2, s_2)
     \end{array} 
   \right)=
   \left(
     \begin{array}{cc}
       1 & 0 \\
       0 & 1 
    \end{array} 
   \right)\mbox{ or }
   \left(
    \begin{array}{cc}
       0 & 1 \\
       1 & 0 
    \end{array} 
   \right). 
\end{equation}


Given an HA-assignment $Y,$ 
  we say that team $t$ has a break at slot $s \in S\setminus \{1\}$
  if $y(t,s-1)=y(t,s).$
The number of breaks in a home–away assignment
  is defined as the total number of breaks belonging to all teams.
For example, in Table~\ref{table:ha_assignment}, 
  team $2$ has a break at slot $2$,
  as there are consecutive away matches $(y(2,1)=y(2,2)=0)$.
Similarly, team $2$ has a break in slot $4$ 
  because of the consecutive home matches  $(y(2,3)=y(2,4)=1)$.
Given a timetable of a round-robin tournament,
  we need to find a consistent HA-assignment
  to complete a schedule. 
In practical sports timetabling
  such as~\cite{nemhauser1998scheduling}, 
  a break is considered undesirable.

\subsection{Previous Studies}

Most methods for constructing a tournament schedule 
  use one of the following two decomposition approaches~\cite{drexl2007sports}:
  (i) first-break-then-schedule and (ii) first-schedule-then-break.
In the first-break-then-schedule approach,
  a matrix of HA-assignment is generated at the first stage.
The second stage finds a timetable consistent with the the fixed HA-assignment, 
  if it exists.
%
In the first-schedule-then-break approach, in contrast,
  a timetable is constructed at the first stage.
The second stage determines the home team and the away team of each match.
In this approach, 
  the home advantage is further determined in the second stage.
In this paper, we address a problem of the second stage 
  in the first-schedule-then-break approach:

\noindent
\textbf{Break Minimization Problem}:
 {\it
 Given a timetable $\tau$, 
 the break minimization problem finds an HA-assignment consistent to $\tau$
 that minimizes the number of breaks.
 }
  
The break minimization problem has been widely studied
  ~\cite{Trick_2001, urdaneta2018alternative, miyashiro2006semidefinite}.
In particular, De Werra~\cite{de1981scheduling} proved that
  the number of breaks in an SRRT schedule is more than $2n - 2$, 
  and the number of breaks in an MDRRT schedule is more than $6n - 6$.
Post and Woeginger~\cite{post2006sports} proved that
  every timetable of an SRRT has an HA-assignment such that
  the number of breaks is
    at most $n(n-1)$ if $n=2k$
    and $(n-1)^2$ if $n = 2k+1$.  
Though the break minimization problems in SRRTs or MDRRTs are known to be hard to solve,
  they have not been proved as NP-hard to the best of our knowledge.


Trick~\cite{Trick_2001} proposed an integer linear programming model
  for the break minimization problem in an SRRT.
The model includes highly symmetry,
  which makes the problem difficult to solve using an integer programming approach.
R{\'e}gin~\cite{regin2000minimization} proposed
  a constraint programming model for the break minimization problem in an SRRT.
Elf, J\"{u}nger and Rinaldi~\cite{ELF2003343} showed
  that solving the break minimization problem in an SRRT
  can be transformed into a maximum cut problem (MAX CUT).  
Miyashiro and Matsui~\cite{miyashiro2006semidefinite} formulated
  the break minimization in an SRRT as a special case of MAX RES CUT
  discussed by Goemans and Williamson~\cite{goemans1995improved};
  thus, it is solvable by an algorithm based on positive semidefinite programming.
Suzuka, et al.~\cite{suzuka2007home}
  addressed the HA-assignment problem and 
  break minimization/maximization problem
  and gave a unified view to the three problems.
Their computational results showed that
  the break minimization problem in an MDRRT schedule is
  more difficult than that in an SRRT schedule.
More precisely, if we solve the problems 
  by an integer programming solver, 
  the average computational time for 
  the break minimization problem in an MDRRT is
  apparently longer than that in an SRRT.
Urdaneta, Yuan and Sinqueria~\cite{urdaneta2018alternative} demonstrated
  a QUBO formulation of the break minimization problem in a DRRT.
They demonstrated that it is superior 
  to the mixed integer linear programming formulation~\cite{Trick_2001}
  for a general mathematical optimization solver.
The QUBO formulation proposed in~\cite{urdaneta2018alternative}
  can also be regarded as an extension of the MAX CUT formulation proposed in~\cite{ELF2003343}.
Kuramata, Katsuki and Nakata~\cite{kuramata2022solving} successfully applied
  quantum annealing to a QUBO formulation.
They used the quantum annealer D-wave Advantage
  and demonstrated that it could obtain better solutions quickly
  compared to a state-of-the-art general solver.

The schedules of most major sports leagues contain 
 restrictions on the number of consecutive home and away matches.
The most common restriction used is an upper bound 
 of consecutive home or away matches~\cite{easton2003solving}.
In practice, the consecutive constraints appear
 as tournament problems of the Danish soccer league~\cite{Rasmussen_2008}.
In this paper, we also address a break minimization problem 
  with {\it consecutive constraints},
  which prohibits consecutive home or away matches
  exceeding two and/or three.
 


\section{Mathematical Formulation}\label{sec:mathform}

In this section, we introduce the QUBO formulation of the break minimization problem
  studied by Urdaneta et al.~\cite{urdaneta2018alternative}.

For each pair $(t,s)\in T \times S$, 
we introduce a binary variable $y_{t,s}\in \{0, 1\}$,
  which takes 1 if team $t$ plays at home in slot $s$ 
  and 0 if it plays away.
A \mbox{0-1} matrix $Y=(y_{t,s})$ is consistent with a given timetable $\tau$ 
  if and only if $Y$ satisfies the following equations: 
\begin{eqnarray}
  \label{eqn:y}
  & y_{t_1, s_1} + y_{t_2, s_1} = 1 \;\;
      (\forall (t_1, t_2, s_1, s_2) \in \Xi (\tau)),  \nonumber \\
  & y_{t_1, s_1} + y_{t_1, s_2} = 1 \;\; 
      (\forall (t_1, t_2, s_1, s_2) \in \Xi (\tau)), \\
  & -y_{t_1, s_1} + y_{t_2, s_2} = 0 \;\; 
      (\forall (t_1, t_2, s_1, s_2) \in \Xi (\tau)).\nonumber
\end{eqnarray}
The above constraints are essentially equivalent to the condition in~(\ref{eqn:xi}).
%
These variables and constraints enable the break minimization problem to be formulated
  as a constrained quadratic binary optimization problem defined as

\begin{align}
  & \text{CQ} : & \text{minimize} & \quad
    \sum_{t \in T} \sum_{s \in S \backslash\{4 n-2\}}
    \Bigl(
        y_{t, s} y_{t, s+1}
        +\left(1-y_{t, s}\right)\left(1-y_{t, s+1}\right)
     \Bigr)  \nonumber \\
  & & \text {subject to} 
      & \quad y_{t_1, s_1} + y_{t_2, s_1} = 1 \quad 
        (\forall (t_1, t_2, s_1, s_2) \in \Xi (\tau)), \label{eqn:redundant1} \\ 
  & & & \quad y_{t_1, s_1} + y_{t_1, s_2} = 1 \quad 
        (\forall (t_1, t_2, s_1, s_2) \in \Xi (\tau)), \label{eqn:redundant2} \\
  & & & \quad -y_{t_1, s_1} + y_{t_2, s_2} = 0 \quad 
        (\forall (t_1, t_2, s_1, s_2) \in \Xi (\tau)), \label{eqn:redundant3} \\
  & & & \quad y_{t, s} \in\{0,1\} \quad(\forall (t,s) \in T \times S). \nonumber
\end{align}
The objective function of CQ consists of the total sums
  of $y_{t, s}y_{t,s+1}$ and $(1 - y_{t, s})(1 - y_{t, s + 1})$,
  which represent the number of breaks corresponding 
  to the two consecutive home and away matches, respectively.

Urdaneta et al.~\cite{urdaneta2018alternative} demonstrated that the constrained optimization problem CQ,
  which represents a break minimization problem in a DRRT,
  can be expressed as an unconstrained quadratic binary optimization problem.
%
%
%
%
%
%
For each $\xi = (t_1, t_2, s_1, s_2) \in \Xi$,
  we introduce a variable $z_{\xi}$ and
  eliminate four variables $y_{t_1, s_1}, y_{t_2, s_1}, y_{t_1, s_2}, y_{t_2, s_2}$
  by the following equalities:
\begin{eqnarray}
  \label{eqn:ztoy}
  y_{t_1, s_1}=z_{\xi}, \quad y_{t_2, s_1} = 1 - z_{\xi}, \quad y_{t_1, s_2}=1 - z_{\xi}, \quad y_{t_2, s_2} = z_{\xi}.
\end{eqnarray}
By substituting equations~(\ref{eqn:ztoy}) into CQ,
  constraint (\ref{eqn:redundant1}) in CQ 
 becomes redundant, 
 as $y_{t_1,s_1}+y_{t_2,s_1}=z_{\xi}+1-z_{\xi}=1.$
In a similar manner, we can show the redundancy of constraints 
 (\ref{eqn:redundant2}) and 
 (\ref{eqn:redundant3}). 
By deleting theses redundant constraints,  
 we obtain an unconstrained optimization problem
  $\{\mbox{\boldmath $z$}^{\top}H\mbox{\boldmath $z$} \mid \mbox{\boldmath $z$} \in \{0, 1\}^{ \Xi }\}$
  for the break minimization problem.

Next, we introduce additional consecutive constraints.
A sequence of consecutive away matches of a team 
is called a {\em road trip}, 
and a sequence of consecutive home matches
is called a {\em home stand}. 
The consecutive constraints for a positive integer $u\geq 2$ is defined as follows:
\begin{description}
\item[\rm CC($u$):] 
the length of any home stand is at most $u$,
and the length of any road trip is at most $u$. 
\end{description}
In this paper, we deal with cases that $u\in \{2,3\}$.
For example, in Table~\ref{table:ha_assignment},
  team 2 plays at away from slot 3 to slot 5,
  which indicates a violation of the constraint CC(2).
The constraint CC(3) has been employed
  in many studies~\cite{easton2003solving,yamaguchi2011improved,miyashiro2012approximation,thielen2011complexity}. 
The constraint CC(2) is discussed in~\cite{thielen2012approximation,imahori2015A1+,xiao2016improved}.

It is well-known that 
  every single round robin timetable has a consistent HA-assignment
  satisfying the constraint CC($u$) for any $u\geq 2$.
For example, 
 the proof of Remark~7 in~\cite{miyashiro2005polynomial}
 gives a procedure to construct an HA-assignment
 (of an SRRT)
in which each team has no breaks at slots $2, 4, \ldots, 2n-2$.
This procedure directly shows that 
 every timetable of MDRRT has an HA-assignment
 satisfying CC(3).
The hardness of the problem of finding
  an HA-assignment satisfying CC(2)
  remains unknown in case of an MDRRT schedule. 

We obtain a formulation of the break minimization problem in MDRRT
with the constraint CC($u$) 
by adding the following linear inequalities to CQ:
%
\begin{equation} \label{eq:consecutive}
 1 \leq y_{t, s} + y_{t, s+1} +\cdots + y_{t, s+u} \leq u \;\;
   (\forall (t,s)\in T \times \{1,2,\ldots , 2n-1\}).
\end{equation}
The first inequality indicates that the number of home matches should be positive,
  which is equivalent to having less than or equal to $u$
  consecutive away matches.
We need not consider cases that $(t,s)\in T \times \{2n,2n+1,\ldots\}$, 
  as  we discuss an MDRRT schedule.
In the case of CC(2), 
  the constraint (\ref{eq:consecutive}) is equivalently expressed 
  in quadratic equality constraints as
\begin{eqnarray}
  \label{eqn:consecutive2}
   (y_{t, s} + y_{t, s+1} + y_{t, s+2} - 1)
  (y_{t, s} + y_{t, s+1} + y_{t, s+2} - 2) = 0  \;\;\; \\ \nonumber
     (\forall (t,s)\in T \times \{1,2,\ldots , 2n-1\}).
\end{eqnarray}
When we address a problem with CC(3),
  we introduce artificial binary variables $w_{t,s}\in \{0,1\}$ 
  and employ the following quadratic equality constraints

\begin{eqnarray}
  \label{eqn:consecutive3}
  && (y_{t, s} + y_{t, s+1} + y_{t, s+2} +y_{t, s+3} +w_{t,s} - 2)
    \\ \nonumber
 &&\;\;\; (y_{t, s} + y_{t, s+1} + y_{t, s+2} +y_{t, s+3} +w_{t,s}- 3) = 0 \\ \nonumber
 && \;\;\;\;\;\;\; \mbox{\qquad}  (\forall (t,s)\in T \times \{1,2,\ldots , 2n-1\}).
\end{eqnarray}

We employ a penalty method to solve the break minimization problem
  with constraints CC(2) and/or CC(3).
We construct a QUBO formulation  
  by adding a penalty function
  (to the objective function)
  that consists of penalty parameters
  multiplied by the left-hand side of~(\ref{eqn:consecutive2}) 
  or~(\ref{eqn:consecutive3}).
It is easy to see that the left-hand sides of both~(\ref{eqn:consecutive2}) 
  and~(\ref{eqn:consecutive3}) are nonnegative
  when the corresponding variables are 0-1 valued. 
Thus, by introducing sufficiently large penalty parameters,
  every optimal solution of the QUBO formulation 
  obtained by the above penalty method
  is also optimal 
  to the break minimization problem with constraints CC(2) and/or CC(3),
  when it is feasible.
Fred Glover et al.~\cite{glover2022quantum} discussed 
  a general technique to incorporate equality constrains 
  in a QUBO formulation. 

\section{Computational Results}\label{sec:comp}

This section presents computational results 
  on break minimization problems in MDRRTs with consecutive constraints.
We report computational results obtained by
  a general MIQP solver, Gurobi Optimizer (version 9.5.2),
  and
  a QUBO solver, QuBowl~\cite{rehfeldt2023faster}.
The computational experiments were performed 
  on Intel (R) Xeon(R) CPU E3-1270 with 32 GB RAM.

QuBowl is a state-of-the-art QUBO solver developed 
  by Rehfeldt, Koch and Shinano~\cite{rehfeldt2023faster}.
It converts QUBO problems to MaxCut problems and
  solve them by conducting a branch-and-cut method.
Enhancing the algorithm with presolving and cutting-plane generation,
  it outperforms state-of-the-art commercial solvers on general QUBO instances.

\subsection*{Testset}

We generate break minimization problem instances
  by the following method from~\cite{kuramata2022solving}.
  
\begin{enumerate}[labelwidth=2cm,leftmargin=1.4cm]
  \item[{\rm Step 1:}] Create an RRT timetable using the Kirkman method~\cite{kirkman1847problem}.
  \item[{\rm Step 2:}] Shuffle the order of the slots in the timetable created in Step~1.
  \item[{\rm Step 3:}] Make a copy of the timetable obtained in Step~2 and concatenate two identical timetables. 
\end{enumerate}

\subsection*{Computational comparison of MIQP and QUBO}

In our experiments, 
  we generated and solved 5 instances for each $n \in \{3,4,\ldots , 15\}$
  using Gurobi and QuBowl.

In the first experiment, we solved (unconstrained) break minimization problems.
Table~\ref{table:MDRRT} provides the results of our first experiment
  for a time-limit of one hour using a single thread.
The 2nd and 3rd columns show the numbers of instances 
  completely solved (within one hour) by Gurobi and QuBowl, respectively.
The 5th and 6th columns show the average execution times of Gurobi and QuBowl,
  where time is set as the time limit when the solvers failed to solve.

Both Gurobi and QuBowl took longer to compute as the number $n$ increased.
Table~\ref{table:MDRRT} clearly shows that QuBowl outperforms Gurobi.
Gurobi failed to solve 3 instances (within one hour) 
  with 28 teams $(n=14)$ and 30 teams $(n=15)$,
  while QuBowl successfully solved all instances.
For 26 teams $(n=13)$,
  QuBowl is faster than Gurobi by a factor of more than 40 on average.

In the second experiment,
  we solved the break minimization problems with consecutive constraints CC(2) and CC(3) using Gurobi and QuBowl.
When we use Gurobi, 
  we solve nonconvex quadratic integer problems
  with linear constraints (\ref{eqn:y}).
If we use QuBowl,
  we penalize quadratic constraints 
  (\ref{eqn:consecutive2}) and/or~(\ref{eqn:consecutive3}) 
  by setting penalty parameters to 10
  to translate them into QUBOs.
Tables~\ref{table:MDRRTwithC2} and~\ref{table:MDRRTwithC3} 
  list the results of the second experiment
  with a time-limit of one hour.
It can be observed that QuBowl
  requires more computational time
  than the first experiment.
We verified that
  the solutions obtained by QuBowl satisfy the constraints CC(2) and/or CC(3). 
Table~\ref{table:MDRRTwithC2} shows that 
 for 26 teams $(n=13)$,
  QuBowl is faster than Gurobi 
  by a factor of more than 15 on average.
Though the running time gaps between QuBowl and Gurobi 
  get close compared with the first experiment,
  QuBowl still outperforms Gurobi for those problems.
In case of CC(3), 
  Table~\ref{table:MDRRTwithC3} shows that 
  QuBowl is competitive with Gurobi.

\begin{table}[hbtp]
  \centering
  \caption{MIQP solver vs. QUBO solver for break minimization problems}      
    \label{table:MDRRT}
    \begin{tabular}{lrrrr}
      \hline &
      \multicolumn{2}{c}{ Solved } & \multicolumn{2}{c}{ run time [s] }  \\
      $n$ & Grb & QuBowl & Grb & QuBowl \\
      \hline
      3 & 5 & 5 & 0.01 & 0.00 \\
      4 & 5 & 5 & 0.03 & 0.01 \\
      5 & 5 & 5 & 0.06 & 0.02 \\
      6 & 5 & 5 & 0.14 & 0.04 \\
      7 & 5 & 5 & 0.57 & 0.10 \\
      8 & 5 & 5 & 1.09 & 0.14 \\
      9 & 5 & 5 & 2.26 & 0.31 \\
      10 & 5 & 5 & 4.98 & 0.76 \\
      11 & 5 & 5 & 36.48 & 2.20 \\
      12 & 5 & 5 & 155.71 & 4.53 \\
      13 & 5 & 5 & 492.31 & 10.43 \\
      14 & 2 & 5 & 2734.31 & 752.48  \\
      15 & 2 & 5 & 2677.92 & 155.88 \\
      \hline
    \end{tabular}
\end{table}

\begin{table}
  \centering
  \caption{MIQP solver vs. QUBO solver for break minimization problems with CC(2).}
\label{table:MDRRTwithC2} 
  \begin{tabular}{lrrrrrrrrr}
    \hline &
    \multicolumn{2}{c}{ Solved } & \multicolumn{2}{c}{ run time [s] } \\
    $n$ & Grb & QuBowl & Grb & QuBowl \\
    \hline
     3 & 5 & 5 & 0.02 & 0.00 \\ 
     4 & 5 & 5 & 0.05 & 0.03 \\ 
     5 & 5 & 5 & 0.08 & 0.05 \\ 
     6 & 5 & 5 & 0.18 & 0.11 \\ 
     7 & 5 & 5 & 0.59 & 0.26 \\ 
     8 & 5 & 5 & 0.93 & 0.28 \\ 
     9 & 5 & 5 & 7.51 & 0.79 \\ 
     10 & 5 & 5 & 20.18 & 4.89 \\ 
     11 & 5 & 5 & 128.43 & 9.33 \\ 
     12 & 5 & 5 & 288.53 & 23.57 \\ 
     13 & 5 & 5 & 687.90 & 61.73 \\ 
     14 & 2 & 5 & 2797.60 & 688.87 \\ 
     15 & 2 & 4 & 2839.15 & 1002.10 \\ 
    \hline
  \end{tabular}
\end{table}  

\begin{table}
  \centering
  \caption{MIQP solver vs. QUBO solver for break minimization problems with CC(3).}
\label{table:MDRRTwithC3} 
  \begin{tabular}{lrrrrrrrrr}
    \hline &
    \multicolumn{2}{c}{ Solved } & \multicolumn{2}{c}{ run time [s] } \\
    $n$ & Grb & QuBowl & Grb & QuBowl \\
    \hline
     3 & 5 & 5 & 0.02 & 0.26 \\ 
     4 & 5 & 5 & 0.04 & 0.39 \\ 
     5 & 5 & 5 & 0.08 & 0.56 \\ 
     6 & 5 & 5 & 0.15 & 0.49 \\ 
     7 & 5 & 5 & 0.27 & 1.35 \\ 
     8 & 5 & 5 & 0.73 & 1.80 \\ 
     9 & 5 & 5 & 3.01 & 5.54 \\ 
     10 & 5 & 5 & 7.87 & 12.32 \\ 
     11 & 5 & 5 & 54.09 & 50.21 \\ 
     12 & 5 & 5 & 165.97 & 107.89 \\ 
     13 & 5 & 5 & 543.06 & 338.84 \\ 
     14 & 3 & 2 & 2523.42 & 2323.49 \\ 
     15 & 3 & 2 & 2403.28 & 2866.82 \\ 
    \hline
  \end{tabular}
\end{table}

\section{Conclusion}\label{sec:concl}

This study demonstrated that
  the QUBO solver outperforms
  a general MIQP solver for break minimization problems.
This indicates an important role of the QUBO solver
  in sport scheduling problems.
Further, the QUBO solver can also outperform 
  in break minimization problems 
  by transforming them into QUBOs.
Although penalizing the quadratic constraints makes the objective function more dense,
  the enhancing techniques in the QUBO solver enables
  more efficient solving than a general MIQP solver.
  
Our results indicate a further potential of the QUBO solver and QUBO formulations.
In recent years, QUBO formulations have been proposed for industry applications, 
  such as nurse scheduling~\cite{ikeda2019application},
  job shop scheduling~\cite{carugno2022evaluating},
  and vehicle routing~\cite{irie2019quantum}.
Although they assumed quantum techniques such as quantum annealing,
  our results suggest that the state-of-the-art QUBO solver could be
  competitive with the general MIQP solver for them.

  \bibliographystyle{elsarticle-num}


\end{document}